\documentclass[aps,pra,twocolumn,showpacs,superscriptaddress]{revtex4-1}

\usepackage[utf8]{inputenc}
\usepackage{amsmath}
\usepackage{amssymb}
\usepackage{latexsym}
\usepackage{ifpdf}

\ifpdf
\usepackage[pdftex]{graphicx}
\else
\usepackage{graphicx}
\fi

\ifpdf
\DeclareGraphicsExtensions{.pdf, .jpg, .tif}
\else
\DeclareGraphicsExtensions{.eps, .jpg}
\fi

\begin{document}

\title{Thermal Casimir force between nanostructured surfaces}

\author{R. Guérout}
\affiliation{Laboratoire Kastler-Brossel, CNRS, ENS, UPMC, Case 74, F-75252 Paris, France}
\author{J. Lussange}
\affiliation{Laboratoire Kastler-Brossel, CNRS, ENS, UPMC, Case 74, F-75252 Paris, France}
\author{H. B. Chan}
\affiliation{Department of Physics, the Hong Kong University of Science and Technology, Hong Kong, China}
\author{A. Lambrecht}
\affiliation{Laboratoire Kastler-Brossel, CNRS, ENS, UPMC, Case 74, F-75252 Paris, France}
\author{S. Reynaud}
\affiliation{Laboratoire Kastler-Brossel, CNRS, ENS, UPMC, Case 74, F-75252 Paris, France}

\date{\today}

\begin{abstract}
We present detailed calculations for the Casimir force between a
plane and a nanostructured surface at finite temperature in the
framework of the scattering theory. We then study numerically the
effect of finite temperature as a function of the grating parameters
and the separation distance. We also infer non-trivial geometrical
effects on the Casimir interaction via a comparison with the proximity
force approximation. Finally, we compare our calculations with data from
experiments performed with nanostructured surfaces.
\end{abstract}


\maketitle

\subsubsection{Introduction}
\label{ssub:introduction}

The Casimir interaction at finite temperature consists of two parts:
a purely quantum one which subsists  as $T\to 0$ K involving the
zero-point energy of the electromagnetic vacuum \cite{casimir1948}
and a thermal one \cite{mehra1967} which takes into account the real
thermal photons emitted by the bodies. For the thermal part to
become noticeable, frequencies relevant to the Casimir interaction
must fall in the range of relevant thermal frequencies. For this
reason, at room temperature the thermal part of the Casimir
interaction becomes important for separation distances of the order
of microns or tens of microns. At those separation distances, the
absolute value of the Casimir force is small as it decreases as the
inverse third power of the separation distance. Experimentally,
there is thus a trade off when one wants to measure the thermal
component of the Casimir interaction: at small separation distance
where the Casimir force is comparably large, the thermal part is
small whereas at large separation distance where the thermal part is
large, the total Casimir force is small. A solution is to use
“large” interacting bodies to maximize the Casimir force at
large distances~\cite{sushkovNatPh2011}.

The above reasoning is well adapted to the parallel-plates
geometry (or a plate-sphere situation when the radius of the sphere
is large) where the only characteristic length is the separation
distance. In the case of a plate-grating situation additional
characteristic lengths such as the grating period or the corrugation
depth are to be taken into account. 
The importance played by the thermal part of the Casimir interaction
is highly non-trivial in the case of the plate-grating geometry and
full calculations become necessary.
% No simple reasoning is possible
% in the plate-grating geometry to estimate the importance of the
% thermal part of the Casimir interaction and full calculations become
% necessary. 
First calculations for this geometry for perfect
reflectors at zero temperature have used a path integral approach
\cite{buscherPRA2004}. A first exact solution for the Casimir force
between two periodic dielectric gratings was given in
\cite{lambrechtPRL2008}. Alternatively, the Casimir force in
such geometries can also be calculated using a
modal approach \cite{davidsPRA2010}. In this paper we use the
scattering approach to Casimir forces
\cite{jaekelJP1991,lambrechtNJP2006,emigPRL2007} to calculate the
Casimir interaction between a plate and a grating at arbitrary
temperature. We study the contribution of the thermal part as a
function of the grating parameters and assess the validity of the
proximity force approximation (PFA). We finally compare our results
with experimental data presented
elsewhere~\cite{chanPRL2008,baoPRL2010}. The most important result
is that in the grating geometry the thermal contribution to the
Casimir interaction is overall enhanced and occurs at shorter
separation distance, which opens interesting perspectives for new
experiments.

\subsubsection{Theory: Thermal Casimir force between gratings}
\label{ssub:theory}

\begin{figure}[htbp]
  \begin{center}
    \includegraphics[width=8cm]{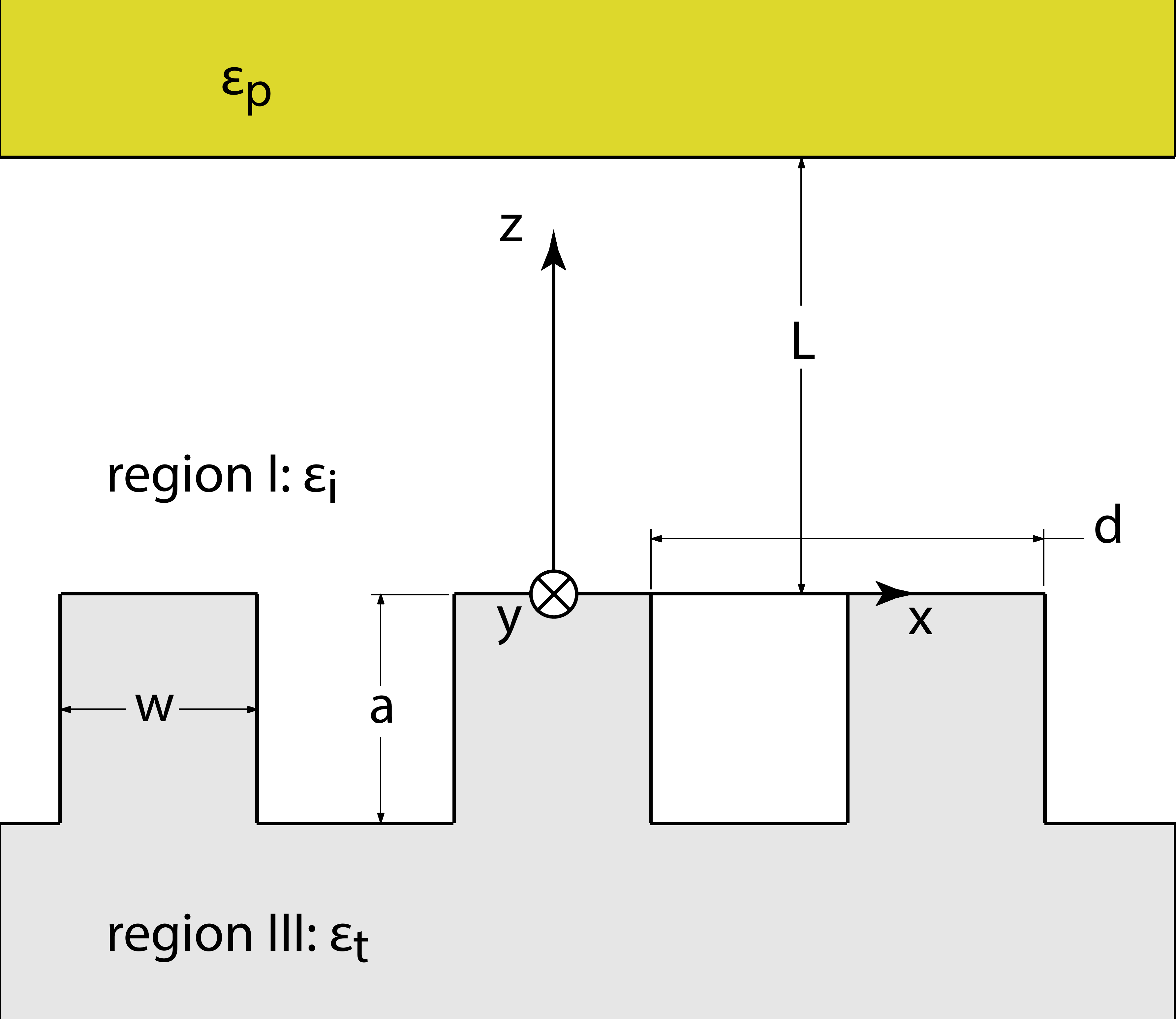}
  \end{center}
  \caption{A 1D lamellar grating and its associated reference frame.}
  \label{fig:grating}
\end{figure}

We study the Casimir interaction within the scattering approach
 between a plate and a 1D lamellar grating as depicted in figure~\ref{fig:grating}. Above
the grating $z>0$, we have a homogeneous region
labelled I characterized by a permittivity $\epsilon_{i}$. Below the
grating $z<-a$, a homogeneous region labeled III
characterized by a permittivity $\epsilon_{t}$. The plate is
characterized by a permittivity $\epsilon_{p}$ for $z>L$. In the
grating region $-a\leq z\leq 0$, the permittivity is a periodic
function of $x$, $\epsilon(x)$. The Casimir force per unit area
$F_{p,g}(L)$ between a plane and a grating separated by a distance
$L$ is calculated in the scattering formalism taking into account
the finite temperature $T$
\begin{equation}
  \begin{split}
      \label{eq:FpgExact}
      F_{p,g}&(L;T)=2\pi k_{B}T\\&\sum_{n=0}^{\infty}{}^{'}\iint\text{tr}\left(\left(\mathbf{1}-\boldsymbol{\mathcal{M}}_{n}\right)^{-1}
      \partial_{L}\boldsymbol{\mathcal{M}}_{n}\right)\mathrm{d}k_{x}\mathrm{d}k_{y}
  \end{split}
\end{equation}
where the prime on the sum means that the term with $n=0$ is to be
multiplied by a factor $1/2$. This expression takes into
account the contribution of a thermal field of real photons whose
mean number $\bar{n}(\omega)=\left(e^{\hbar\omega/k_{B}
T}-1\right)^{-1}$ follows a Planck's law to the contribution of the
virtual photons emerging from the electromagnetic vacuum. In the
following we will use $c=1$ for convenience and work with a generalized 
complex frequency $\Omega=\omega+\imath \xi$ having real and imaginary
parts $\omega$ and $\xi$ respectively.

The function $\boldsymbol{\mathcal{M}}_{n}$  is evaluated at the
Matsubara frequency $\xi_{n}=\frac{2\pi n k_{B}T}{\hbar}$ and reads
$\boldsymbol{\mathcal{M}}(\imath\xi_{n})=\mathbf{R}_{p}(\imath\xi_{n})e^{-\boldsymbol{\kappa}
L}\mathbf{R}_{g}(\imath\xi_{n})e^{-\boldsymbol{\kappa} L}$. It
describes a full round-trip of the field between the two scatterers,
that is a reflection on the plate via the operator $\mathbf{R}_{p}$,
the free propagation from the plate to the grating corresponding to
the translational operator $e^{-\boldsymbol{\kappa} L}$  with the
imaginary wave vector $\boldsymbol{\kappa}=-\imath
\mathbf{k}_{z}=\left(\epsilon_{i}\xi_{n}^{2}+\mathbf{k}_{x}^{2}+k_{y}^{2}\right)^{1/2}$,
the scattering on the grating via the reflection operator
$\mathbf{R}_{g}$ and a free propagation back to the plate. The
vector $\mathbf{k}_{x}$ of dimension $2N+1$ collects the
diffracted wavevectors
$\mathbf{k}_{x}=(k_{x}-N\frac{2\pi}{d},\cdots,k_{x},\cdots,k_{x}+N\frac{2\pi}{d})$
where $N$ is the highest diffraction order retained in the calculation.
The integration in eq.~(\ref{eq:FpgExact}) is restricted to the
first Brillouin zone \emph{i.e.} $-\frac{\pi}{d}\leq k_{x}\leq
\frac{\pi}{d}$.

The plate's reflection operator $\mathbf{R}_{p}$ is diagonal and
collects the appropriate Fresnel reflection coefficients 
$\mathbf{R}_{p}=\text{diag}(\frac{\boldsymbol{\kappa}-\boldsymbol{\kappa}_{p}}{\boldsymbol{\kappa}+\boldsymbol{\kappa}_{p}},
\frac{\epsilon_{p}\boldsymbol{\kappa}-\epsilon_{i}\boldsymbol{\kappa}_{p}}{\epsilon_{p}\boldsymbol{\kappa}+\epsilon_{i}\boldsymbol{\kappa}_{p}})$
with
$\boldsymbol{\kappa}_{p}=\left(\epsilon_{p}\xi_{n}^{2}+\mathbf{k}_{x}^{2}+k_{y}^{2}\right)^{1/2}$.
We calculate the grating reflection operator $\mathbf{R}_{g}$ in the
framework of the Rigorous Coupled Wave Analysis (RCWA). We use a
method of resolution inspired by the formalism presented \emph{e.g.}
in~\cite{moharam1995} which we will briefly outline below.

The physical problem is time- and $y$-invariant. A global dependence in $e^{\imath(k_{y} y-\omega t)}$ can be factored out of all the fields.

For an incident wave characterized by a wave vector
$\mathbf{k}_{i}^{(p)}=(\alpha_{p},k_{y},-\gamma_{p}^{(i)})$ a
grating structure with period $d$ will generate an infinite number
of reflected waves with wave vectors
$\mathbf{k}_{r}^{(n)}=(\alpha_{n},k_{y},\gamma_{n}^{(i)})$ and
transmitted waves with wave vectors
$\mathbf{k}_{t}^{(n)}=(\alpha_{n},k_{y},-\gamma_{n}^{(t)})$.
$\alpha_{n}=k_{x}+n\frac{2\pi}{d}$ and
$\gamma_{n}^{(i/t)}=\left(\epsilon_{i/t}\omega^{2}-\alpha_{n}^{2}-k_{y}^{2}\right)^{1/2}$.

In region I, the $y$-components of the fields are written as a
Rayleigh expansion involving incident and reflected fields of order
$p$ and $n$ respectively
\begin{subequations}
    \label{eq:rayleighI}
    \begin{align}
      &\begin{split}
          E_{y}(x,z)&=\sum_{n=-\infty}^{+\infty}\delta_{np}I_{p}^{(e)}e^{\imath(\alpha_{p}x-\gamma_{p}^{(i)}z)}\\
          &+ R_{np}^{(e)} e^{\imath(\alpha_{n}x+\gamma_{n}^{(i)}z)}
      \end{split}\\
      &\begin{split}
          H_{y}(x,z)&=\sum_{n=-\infty}^{+\infty}\delta_{np}I_{p}^{(h)}e^{\imath(\alpha_{p}x-\gamma_{p}^{(i)}z)}\\
          &+ R_{np}^{(h)} e^{\imath(\alpha_{n}x+\gamma_{n}^{(i)}z)}
      \end{split}
    \end{align}
\end{subequations}
whereas in region III, the Rayleigh expansion involves the
transmitted fields
\begin{subequations}
    \label{eq:rayleighIII}
    \begin{align}
      E_{y}(x,z)&=\sum_{n=-\infty}^{+\infty} T_{np}^{(e)} e^{\imath(\alpha_{n}x-\gamma_{n}^{(t)}(z+a))}\\
      H_{y}(x,z)&=\sum_{n=-\infty}^{+\infty} T_{np}^{(h)} e^{\imath(\alpha_{n}x-\gamma_{n}^{(t)}(z+a))}
    \end{align}
\end{subequations}

Whether in region I or III, the $x$-components of the fields are obtained from the $y$-components thanks to the Maxwell's curl equations
\begin{subequations}
    \label{eq:maxwell}
    \begin{align}
      &\begin{split}
          E_{x}(x,z)&=\frac{\imath k_{y}}{\epsilon_{i/t}\omega^{2}-k_{y}^{2}}\partial_{x}E_{y}(x,z)\\
          &+\frac{\imath \omega}{\epsilon_{i/t}\omega^{2}-k_{y}^{2}}\partial_{z}H_{y}(x,z)
      \end{split}\\
      &\begin{split}
          H_{x}(x,z)&=\frac{\imath k_{y}}{\epsilon_{i/t}\omega^{2}-k_{y}^{2}}\partial_{x}H_{y}(x,z)\\
          &-\frac{\imath \omega\epsilon_{i/t}}{\epsilon_{i/t}\omega^{2}-k_{y}^{2}}\partial_{z}E_{y}(x,z)
      \end{split}
    \end{align}
\end{subequations}

In the grating region $-a\leq z\leq 0$, owing to the periodicity
along the $x$ direction the fields  as well as the permittivity
$\epsilon(x)$ and its reciprocal $1/\epsilon(x)$ can be expanded in
Fourier series. We have
\begin{subequations}
    \begin{align}
      \label{eq:fourier}
      E_{x}(x,z)&=\sum_{n}e_{x}^{(n)}(z)e^{\imath k_{x}x}e^{2\imath n \pi x/d}\\
      E_{y}(x,z)&=\sum_{n}e_{y}^{(n)}(z)e^{\imath k_{x}x}e^{2\imath n \pi x/d}\\
      H_{x}(x,z)&=\sum_{n}h_{x}^{(n)}(z)e^{\imath k_{x}x}e^{2\imath n \pi x/d}\\
      H_{y}(x,z)&=\sum_{n}h_{y}^{(n)}(z)e^{\imath k_{x}x}e^{2\imath n \pi x/d}\\
      \epsilon(x)&=\sum_{n}\epsilon_{n}e^{2\imath n \pi x/d}\\
      1/\epsilon(x)&=\sum_{n}\backepsilon_{n}e^{2\imath n \pi x/d}
    \end{align}
\end{subequations}

With those notations, we are able to express the Maxwell's curl
equations in a compact matrix form. Let $\mathbf{F}$ be  a column
vector collecting the Fourier components of the fields
$\mathbf{F}=(\mathbf{e}_{x},\mathbf{e}_{y},\mathbf{h}_{x},\mathbf{h}_{y})^{\mathsf{T}}$,
we may write
\begin{widetext}
    \begin{subequations}
        \begin{align}
          \label{eq:matrixM1}
          \partial_{z}\mathbf{F}&=\mathbf{M}\mathbf{F}\\
          \label{eq:matrixM2}
          \mathbf{M}&=\begin{pmatrix}
            \mathbf{0} & \mathbf{0} & -\frac{\imath k_{y}}{\omega}\boldsymbol{\alpha}\boldsymbol{\epsilon}^{-1} & -\imath\omega \mathbf{1}+\frac{\imath}{\omega}\boldsymbol{\alpha}\boldsymbol{\epsilon}^{-1}\boldsymbol{\alpha} \\
            \mathbf{0} & \mathbf{0} & \imath\omega \mathbf{1}-\frac{\imath k_{y}^{2}}{\omega}\boldsymbol{\epsilon}^{-1} & \frac{\imath k_{y}}{\omega}\boldsymbol{\epsilon}^{-1}\boldsymbol{\alpha}\\
            \frac{\imath k_{y}}{\omega}\boldsymbol{\alpha} & \imath\omega \boldsymbol{\epsilon}-\frac{\imath}{\omega}\boldsymbol{\alpha}\boldsymbol{\alpha} & \mathbf{0} & \mathbf{0} \\
            -\imath\omega \pmb{\backepsilon}^{-1}+\frac{\imath k_{y}^{2}}{\omega} \mathbf{1} & -\frac{\imath k_{y}}{\omega}\boldsymbol{\alpha} & \mathbf{0} & \mathbf{0}
          \end{pmatrix}\equiv \begin{pmatrix}
          \mathbf{0} & \mathbf{M}_{1}\\
          \mathbf{M}_{2} & \mathbf{0}
          \end{pmatrix}
        \end{align}
    \end{subequations}
\end{widetext}

In the above equation,
$\boldsymbol{\alpha}=\text{diag}(\alpha_{n})$, $\mathbf{1}$ is the
identity and $\boldsymbol{\epsilon}$, resp. $\pmb{\backepsilon}$,
are Toeplitz matrices whose structure is defined as having elements
$\{\epsilon_{n},n\geq0\}$ on the first line and elements
$\{\epsilon_{n},n\leq0\}$ on the first column. Note that in
accordance with ref.~\cite{lalanneJOSA1996}, we have replaced the
matrix $\boldsymbol{\epsilon}$ by $\pmb{\backepsilon}^{-1}$ in the
lower-left block of the matrix $\mathbf{M}$. This constitutes an
improvement with respect to ref. \cite{lambrechtPRL2008}, where this
replacement had not been done. The matrix $\mathbf{F}$ has dimension
$4\times 1$ in units of $2N+1$. A particular column of $\mathbf{F}$
corresponds to a particular incident order $p$. In the following,
bold quantities are matrices whose dimensions will be given in units
of $2N+1$ if not trivial. As an example, in eq.~(\ref{eq:matrixM2})
$\boldsymbol{\alpha}$, $\boldsymbol{\epsilon}$, $\pmb{\backepsilon}$
and $\mathbf{1}$ are matrices of dimension $1\times 1$ whereas
$\mathbf{M}_{1}$ and $\mathbf{M}_{2}$ are of dimension $2\times 2$.

In \cite{lambrechtPRL2008} eqn. (\ref{eq:matrixM1}) had been
numerically solved. Here we follow a different path which has proven
to lead to more stable numerical calculations. Because of the block
anti-diagonal structure of matrix $\mathbf{M}$,
eq.~(\ref{eq:matrixM1})  can be recast as a Helmholtz-like equation
for the electric fields provided that $\mathbf{M}$ is
independent of $z$ which is the case for the lamellar gratings we
consider here
\begin{equation}
  \label{eq:helm}
  \partial^{2}_{z^{2}}\begin{pmatrix}
  \mathbf{e}_{x} \\
  \mathbf{e}_{y}
  \end{pmatrix}=\mathbf{M}_{1}\mathbf{M}_{2}\begin{pmatrix}
  \mathbf{e}_{x} \\
  \mathbf{e}_{y}
  \end{pmatrix}\equiv \mathbf{M}^{(e)}\begin{pmatrix}
  \mathbf{e}_{x} \\
  \mathbf{e}_{y}
  \end{pmatrix}
\end{equation}

This equation is solved as $\begin{pmatrix}\mathbf{e}_{x}
\\\mathbf{e}_{y}\end{pmatrix}(z)=e^{\sqrt{\mathbf{M}^{(e)}}z}\mathbf{C^{+}}+e^{-\sqrt{\mathbf{M}^{(e)}}z}\mathbf{C^{-}}$
where $\mathbf{C^{+}}$ and $\mathbf{C^{-}}$ are unknown coefficients
to be determined. Let $\boldsymbol{\phi}$, $\boldsymbol{\lambda}$ be
respectively the eigenvectors and eigenvalues of the matrix
$\mathbf{M}^{(e)}$ such that
$\mathbf{M}^{(e)}=\boldsymbol{\phi}\,\text{diag}(\boldsymbol{\lambda})\,\boldsymbol{\phi}^{-1}$.
Writing explicitly the expression for $e^{\pm
\sqrt{\mathbf{M}^{(e)}}z}$, we can include the matrix
$\boldsymbol{\phi}^{-1}$ in the unknown coefficients
$\mathbf{C^{+}}$ and $\mathbf{C^{-}}$; furthermore we want to avoid
exponentially growing solutions at $z=-a$. Following the
prescriptions in~\cite{moharam1995} we finally arrive at
\begin{subequations}
  \label{eq:helmSolution}
    \begin{align}
      \label{eq:helmSolution1}
      \begin{pmatrix}\mathbf{e}_{x} \\\mathbf{e}_{y}\end{pmatrix}(z)&=\boldsymbol{\phi}e^{\sqrt{\boldsymbol{\lambda}}z}\mathbf{C^{+}}+\boldsymbol{\phi}e^{-\sqrt{\boldsymbol{\lambda}}(z+a)}\mathbf{C^{-}}\\
        \label{eq:helmSolution2}
      \begin{split}
          \begin{pmatrix}\mathbf{h}_{x} \\\mathbf{h}_{y}\end{pmatrix}(z)&=\mathbf{M}^{-1}_{1}\boldsymbol{\phi}\sqrt{\boldsymbol{\lambda}}e^{\sqrt{\boldsymbol{\lambda}}z}\mathbf{C^{+}}\\&-\mathbf{M}^{-1}_{1}\boldsymbol{\phi}\sqrt{\boldsymbol{\lambda}}e^{-\sqrt{\boldsymbol{\lambda}}(z+a)}\mathbf{C^{-}}
      \end{split}
    \end{align}
\end{subequations}
where eq.~(\ref{eq:helmSolution2}) has been obtained by injecting eq.~(\ref{eq:helmSolution1}) into $\begin{pmatrix}\mathbf{h}_{x} \\ \mathbf{h}_{y}\end{pmatrix}=\mathbf{M}^{-1}_{1}\partial_{z}\begin{pmatrix}\mathbf{e}_{x} \\ \mathbf{e}_{y}\end{pmatrix}$ from eq.~(\ref{eq:matrixM2}).

We can use eqs.~(\ref{eq:rayleighI}) and eqs.~(\ref{eq:maxwell}) to
write the fields at $z=0$ and eqs.~(\ref{eq:rayleighIII})  to write
the fields at $z=-a$. In compact matrix form, this leads to
    \begin{align}
    \label{eq:fieldsa}
      \begin{split}
          \mathbf{F}(-a)&=
          \begin{pmatrix}
            -\frac{k_{y}\boldsymbol{\alpha}}{\epsilon_{t}\omega^{2}-k_{y}^{2}} & \frac{\omega\boldsymbol{\gamma}^{(t)}}{\epsilon_{t}\omega^{2}-k_{y}^{2}} \\
            \mathbf{1} & \mathbf{0} \\
            -\frac{\omega\epsilon_{t}\boldsymbol{\gamma}^{(t)}}{\epsilon_{t}\omega^{2}-k_{y}^{2}} & -\frac{k_{y}\boldsymbol{\alpha}}{\epsilon_{t}\omega^{2}-k_{y}^{2}} &  \\
            \mathbf{0} & \mathbf{1}
          \end{pmatrix}
          \begin{pmatrix}
            \mathbf{T}^{(e)} \\
            \mathbf{T}^{(h)}
          \end{pmatrix}\\&\equiv  \begin{pmatrix} \mathbf{t}_{e} \\ \mathbf{t}_{h}\end{pmatrix} \begin{pmatrix}
            \mathbf{T}^{(e)} \\
            \mathbf{T}^{(h)}
          \end{pmatrix}
      \end{split}
    \end{align}
and
    \begin{align}
    \label{eq:fields0}
      \begin{split}
          \mathbf{F}(0)&=
          \begin{pmatrix}
            -\frac{k_{y}\boldsymbol{\alpha}}{\epsilon_{i}\omega^{2}-k_{y}^{2}} & \frac{\omega\boldsymbol{\gamma}^{(i)}}{\epsilon_{i}\omega^{2}-k_{y}^{2}}\\
            \mathbf{1} & \mathbf{0} \\
            -\frac{\omega\epsilon_{i}\boldsymbol{\gamma}^{(i)}}{\epsilon_{i}\omega^{2}-k_{y}^{2}} & -\frac{k_{y}\boldsymbol{\alpha}}{\epsilon_{i}\omega^{2}-k_{y}^{2}}\\
            \mathbf{0} & \mathbf{1}
          \end{pmatrix}
          \begin{pmatrix}
          \mathbf{I}^{\sigma=e} \\
          \mathbf{I}^{\sigma=h}
        \end{pmatrix} \\
          &+\begin{pmatrix}
            -\frac{k_{y}\boldsymbol{\alpha}}{\epsilon_{i}\omega^{2}-k_{y}^{2}} & -\frac{\omega\boldsymbol{\gamma}^{(i)}}{\epsilon_{i}\omega^{2}-k_{y}^{2}} \\
            \mathbf{1} & \mathbf{0} \\
            \frac{\omega\epsilon_{i}\boldsymbol{\gamma}^{(i)}}{\epsilon_{i}\omega^{2}-k_{y}^{2}} & -\frac{k_{y}\boldsymbol{\alpha}}{\epsilon_{i}\omega^{2}-k_{y}^{2}} &  \\
            \mathbf{0} & \mathbf{1}
          \end{pmatrix}
          \begin{pmatrix}
            \mathbf{R}^{(e)} \\
            \mathbf{R}^{(h)}
          \end{pmatrix}\\ &\equiv
          \begin{pmatrix}
            \mathbf{i}_{ee} & \mathbf{i}_{eh} \\
            \mathbf{i}_{he} & \mathbf{i}_{hh}
        \end{pmatrix}  \begin{pmatrix}
          \mathbf{I}^{\sigma=e} \\
          \mathbf{I}^{\sigma=h}
          \end{pmatrix}+\begin{pmatrix} \mathbf{r}_{e} \\ \mathbf{r}_{h}\end{pmatrix} \begin{pmatrix}
                \mathbf{R}^{(e)} \\
                \mathbf{R}^{(h)}
              \end{pmatrix}
      \end{split}
    \end{align}
where $\boldsymbol{\gamma}^{(i/t)}=\text{diag}(\gamma_{n}^{(i/t)})$
and we have introduced the basis of polarizations $\sigma$ we use,
denoted $e$ and $h$. The polarizations $\sigma=e,h$ are defined by
imposing $H_{y}=0$ and $E_{y}=0$ respectively. Hence, in the above
equation for incident $\sigma=e$ waves, we impose
$\mathbf{I}^{\sigma=e}=\mathbf{1}$ and
$\mathbf{I}^{\sigma=h}=\mathbf{0}$ and vice-versa for incident
$\sigma=h$ waves. Note that $\mathbf{t}_{e}$, $\mathbf{t}_{h}$,
$\mathbf{r}_{e}$ and $\mathbf{r}_{h}$ are of dimension $2\times 2$
whereas $\mathbf{i}_{ee}$, $\mathbf{i}_{eh}$, $\mathbf{i}_{he}$ and
$\mathbf{i}_{hh}$ are of dimension $2\times 1$. Other dimensions are
deduced so as to be consistent with those of $\mathbf{F}$.

Evaluating eqs.~(\ref{eq:helmSolution}) at $z=-a$ and $z=0$ and
identifying with eqs.~(\ref{eq:fieldsa}) and ~(\ref{eq:fields0})
leads to a linear system of equations of dimension $8(2N+1)$ for the
$8(2N+1)$ unknowns $\mathbf{C^{+}}$, $\mathbf{C^{-}}$,
$\mathbf{R}^{(e)}$, $\mathbf{R}^{(h)}$, $\mathbf{T}^{(e)}$ and
$\mathbf{T}^{(h)}$. Nevertheless, it is numerically more stable to
eliminate the reflection and transmission unknowns from this system
and to solve instead a reduced system of dimension $4(2N+1)$ for
solely $\mathbf{C^{+}}$ and $\mathbf{C^{-}}$. All done, this system
reads:

\begin{widetext}
    \begin{equation}
      \label{eq:cCoeffs}
      \begin{pmatrix}
        \left(\boldsymbol{\phi}-\mathbf{t}_{e}\mathbf{t}^{-1}_{h}\mathbf{V}\right)e^{-\sqrt{\boldsymbol{\lambda}}a} & \boldsymbol{\phi}+\mathbf{t}_{e}\mathbf{t}^{-1}_{h}\mathbf{V} \\
        \boldsymbol{\phi}-\mathbf{r}_{e}\mathbf{r}^{-1}_{h}\mathbf{V} & \left(\boldsymbol{\phi}+\mathbf{r}_{e}\mathbf{r}^{-1}_{h}\mathbf{V}\right)e^{-\sqrt{\boldsymbol{\lambda}}a}
      \end{pmatrix}
      \begin{pmatrix}
        \mathbf{C^{+}} \\
        \mathbf{C^{-}}
    \end{pmatrix}=
    \begin{pmatrix}
      \mathbf{0} \\
      \begin{pmatrix}\mathbf{i}_{ee}-\mathbf{r}_{e}\mathbf{r}^{-1}_{h}\mathbf{i}_{he} & \mathbf{i}_{eh}-\mathbf{r}_{e}\mathbf{r}^{-1}_{h}\mathbf{i}_{hh}
      \end{pmatrix}  \begin{pmatrix}
          \mathbf{I}^{\sigma=e} \\
          \mathbf{I}^{\sigma=h}
          \end{pmatrix}
    \end{pmatrix}
    \end{equation}
\end{widetext}
where we have defined
$\mathbf{V}=\mathbf{M}^{-1}_{1}\boldsymbol{\phi}\sqrt{\boldsymbol{\lambda}}$.
Once the unknown coefficients $\mathbf{C^{+}}$ and $\mathbf{C^{-}}$
are determined by solving eq.~(\ref{eq:cCoeffs}), the reflection and
transmission coefficients are:
\begin{subequations}
    \label{eq:RTCoeffs}
    \begin{align}
      \begin{pmatrix}
        \mathbf{R}^{(e)} \\
        \mathbf{R}^{(h)}
      \end{pmatrix}
      &=\mathbf{r}^{-1}_{h}\left(\mathbf{V}\left(\mathbf{C^{+}}-e^{-\sqrt{\boldsymbol{\lambda}}a}\mathbf{C^{-}}\right)\right. \\ \nonumber & \left. -\begin{pmatrix}
      \mathbf{i}_{he} & \mathbf{i}_{hh}
  \end{pmatrix}
  \begin{pmatrix}
    \mathbf{I}^{\sigma=e} \\
    \mathbf{I}^{\sigma=h}
  \end{pmatrix}
        \right)\\
      \begin{pmatrix}
        \mathbf{T}^{(e)} \\
        \mathbf{T}^{(h)}
      \end{pmatrix}
      &=\mathbf{t}^{-1}_{h}\left(\mathbf{V}\left(e^{-\sqrt{\boldsymbol{\lambda}}a}\mathbf{C^{+}}-\mathbf{C^{-}}\right)\right)
    \end{align}
\end{subequations}

Resolution of eq.~(\ref{eq:cCoeffs}) and eqs.~(\ref{eq:RTCoeffs})
first for incident $\sigma=e$ waves and then for incident $\sigma=h$
waves leads to the complete reflection and transmission matrices
$\mathbf{R}_{g}$ and $\mathbf{T}_{g}$:
\begin{subequations}
    \label{eq:RTmatrices}
    \begin{align}
      \mathbf{R}_{g}&=
      \begin{pmatrix}
        \mathbf{R}^{(e)}(\sigma=e) & \mathbf{R}^{(e)}(\sigma=h) \\
        \mathbf{R}^{(h)}(\sigma=e) & \mathbf{R}^{(h)}(\sigma=h)
      \end{pmatrix} \\
      \mathbf{T}_{g}&=
      \begin{pmatrix}
        \mathbf{T}^{(e)}(\sigma=e) & \mathbf{T}^{(e)}(\sigma=h) \\
        \mathbf{T}^{(h)}(\sigma=e) & \mathbf{T}^{(h)}(\sigma=h)
      \end{pmatrix}
    \end{align}
\end{subequations}
The force at $T=0$ K is recovered by the substitution:
\begin{equation}
  \label{eq:subs}
  2\pi k_{B}T\sum_{n=0}^{\infty}{}^{'}\to\hbar\int_{0}^{\infty}\mathrm{d}\xi
\end{equation}

We define two quantities, $\vartheta_{F}(L)$ and $\eta_{F}(L)$ to
assess respectively the effect of the finite temperature and the
deviation from PFA:
\begin{align}
  \label{eq:thetaEta}
  \vartheta_{F}(L)&=\frac{F_{p,g}(L;T=300\,K)}{F_{p,g}(L;T=0\,K)}\\
  \eta_{F}(L;T)&=\frac{F_{p,g}(L;T)}{F^{PFA}_{p,g}(L;T)}
\end{align}
where
$F^{PFA}_{p,g}(L;T)=\frac{1}{d}\int_{0}^{d}F_{p,p}(L(x);T)\text{d}x$
and $F_{p,p}(L;T)$ the force  between two plane-parallel plates is
given by the Lifshitz formula~\cite{lifshitz1956}.

\subsubsection{Numerical evaluations}
\label{ssub:applications}

We now study the thermal Casimir interaction between a gold plate
and a doped silicon grating. Therefore,
$\epsilon_{t}\equiv\epsilon_{Si}(\omega)$ and
$\epsilon_{p}\equiv\epsilon_{Au}(\omega)$. The plate and the grating
are separated by vacuum
$\epsilon_{i}\equiv\epsilon_{0}=1$ $\forall \omega$. As described
in~\cite{baoPRL2010}, the permittivity of gold is taken from
experimental data extrapolated to low frequencies by the Drude model.
The permittivity of doped silicon is modeled by a two-oscillator
model: one describing the intrinsic part of silicon and the other
one describing its metallic behavior at low frequencies
\cite{lambrechtEPL2007}. The metallic part of doped silicon is
determined by a doping level of $2\times 10^{18}$
cm$^{-3}$. The situation is characterized by four length scales: the
separation distance $L$, the corrugation height $a$, the grating
period $d$ and the corrugation width $w$ (for this last quantity, we
will prefer to work with the filling factor $f=w/d$). A complete
analysis would in principle involve full Casimir force
calculations in a four dimensional parameters space. Instead we
explore here the parameter space at fixed filling factor and grating
period $d$ corresponding to the experimental set-up
in~\cite{baoPRL2010}.

\begin{figure}[htbp]
  \begin{center}
    \includegraphics[width=3in]{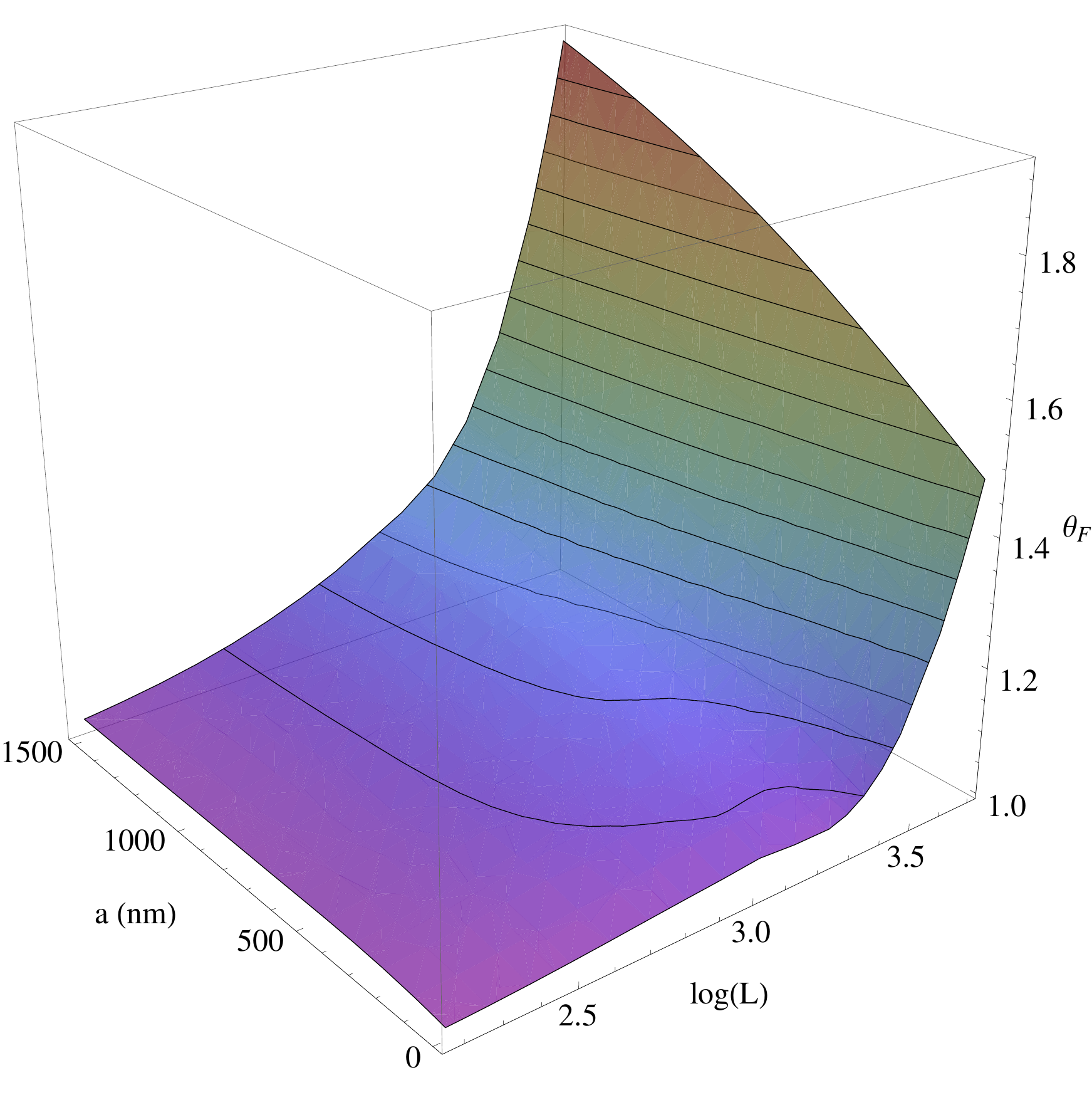}
  \end{center}
  \caption{Effect of temperature on the Casimir force given by the ratio $\vartheta_{F}$ as a function of the separation distance $L$ and corrugation depth $a$.}
  \label{fig:thetaF}
\end{figure}

In figure~\ref{fig:thetaF}
we illustrate the effect of the temperature in the plate-grating
geometry by plotting the ratio $\vartheta_{F}(L,a)$ as a function of
both the separation distance $L$ and the trench depth $a$. The
grating period $d$ and the filling factor $f$ are fixed respectively
at $d=400$ nm and $f=0.5$. For this choice of materials, we find over
the whole range of parameters $\vartheta_{F}(L,a)>1$ so that the
thermal photons always lead to an increase in the Casimir force. For
$a=0$ we recover the two-plate configuration and we have
necessarily $\lim_{L \to 0}\vartheta_{F}(L,a=0)=1$. The total
temperature effect $\vartheta_{F}(L,a=\infty)-\vartheta_{F}(L,a=0)$
increases with larger separation distances $L$. The limiting
value $\vartheta_{F}(L,a=\infty)$ is reached for larger $a$ as the
separation distance $L$ increases since this limiting value rather
means $a>>L$.

\begin{figure}[htbp]
  \begin{center}
    \includegraphics[width=3in]{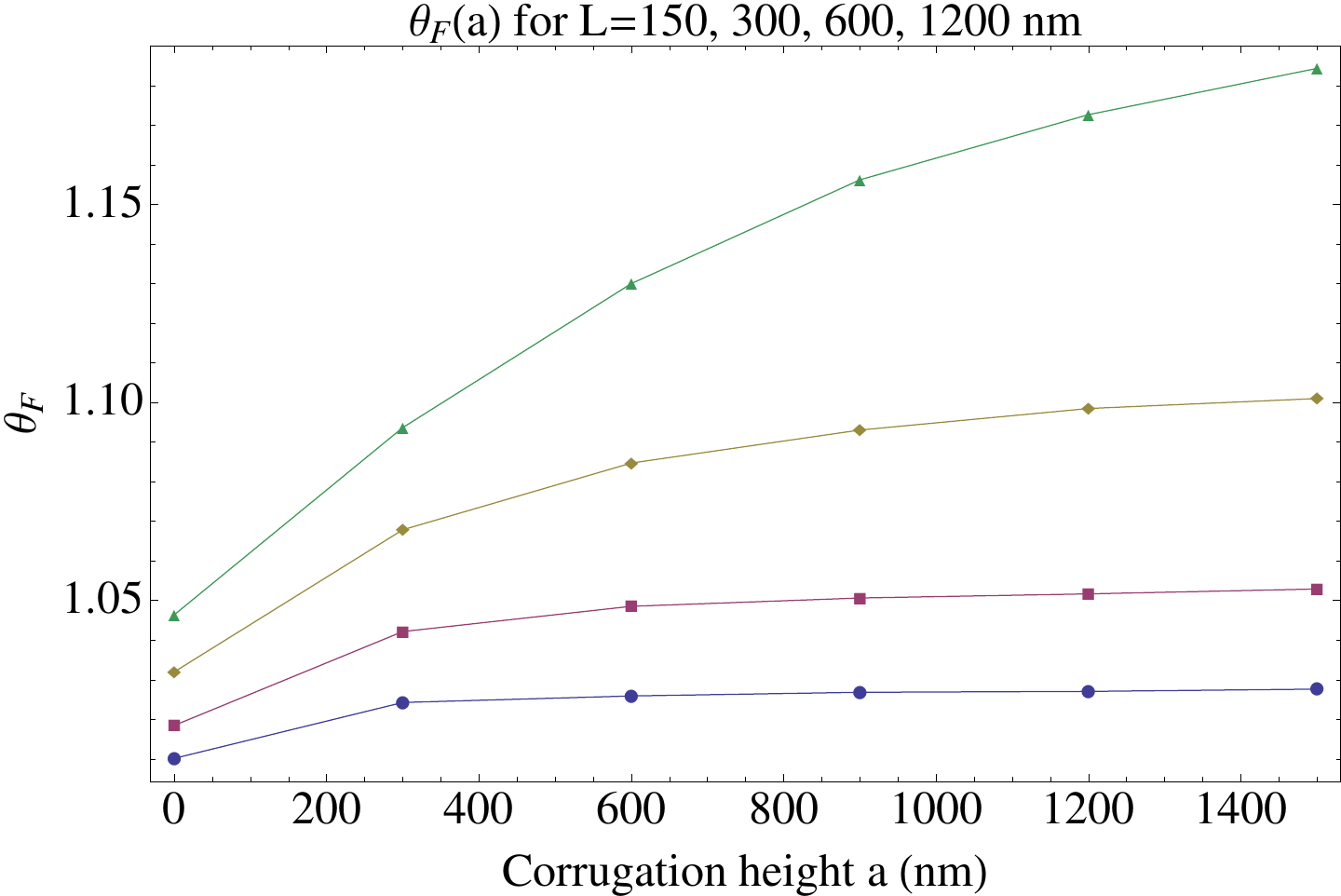}
  \end{center}
  \caption{Effect of the temperature $\vartheta_{F}$ on the Casimir force for $L$=150, 300, 600, 1200 nm (from bottom to top) as a function of the trench depth $a$.}
  \label{fig:thetaF_cut}
\end{figure}

Interestingly, for a fixed separation distance $L$, there is a steep
increase of $\vartheta_{F}$ as a function of the corrugation depth
$a$ towards saturation as shown in detail in
figure~\ref{fig:thetaF_cut}.
At a distance of $1.2\mu$m the temperature corrections increase the zero
temperature force by $\sim 5\%$ between two flat plates. Remarkably,
this increase becomes $\sim
20\%$ if the doped Si plate contains deep trenches ($a\sim
1.4\mu$m). For $L=600$ nm the effect is less pronounced but still
amounts to an increase from 3\% to about 10\%. Clearly the use of a
structured surface increases the thermal Casimir force and makes the
effect easier to be observed at shorter distances. A possible explanation is
that the nanostructures change the spectral mode density especially
in the infrared frequency domain, so that thermal effects become
enhanced, as it has already been pointed out in heat
transfer phenomena between gratings
\cite{gueroutPRB2012,lussangePRB2012}.

\begin{figure}[htbp]
  \begin{center}
    \includegraphics[width=3in]{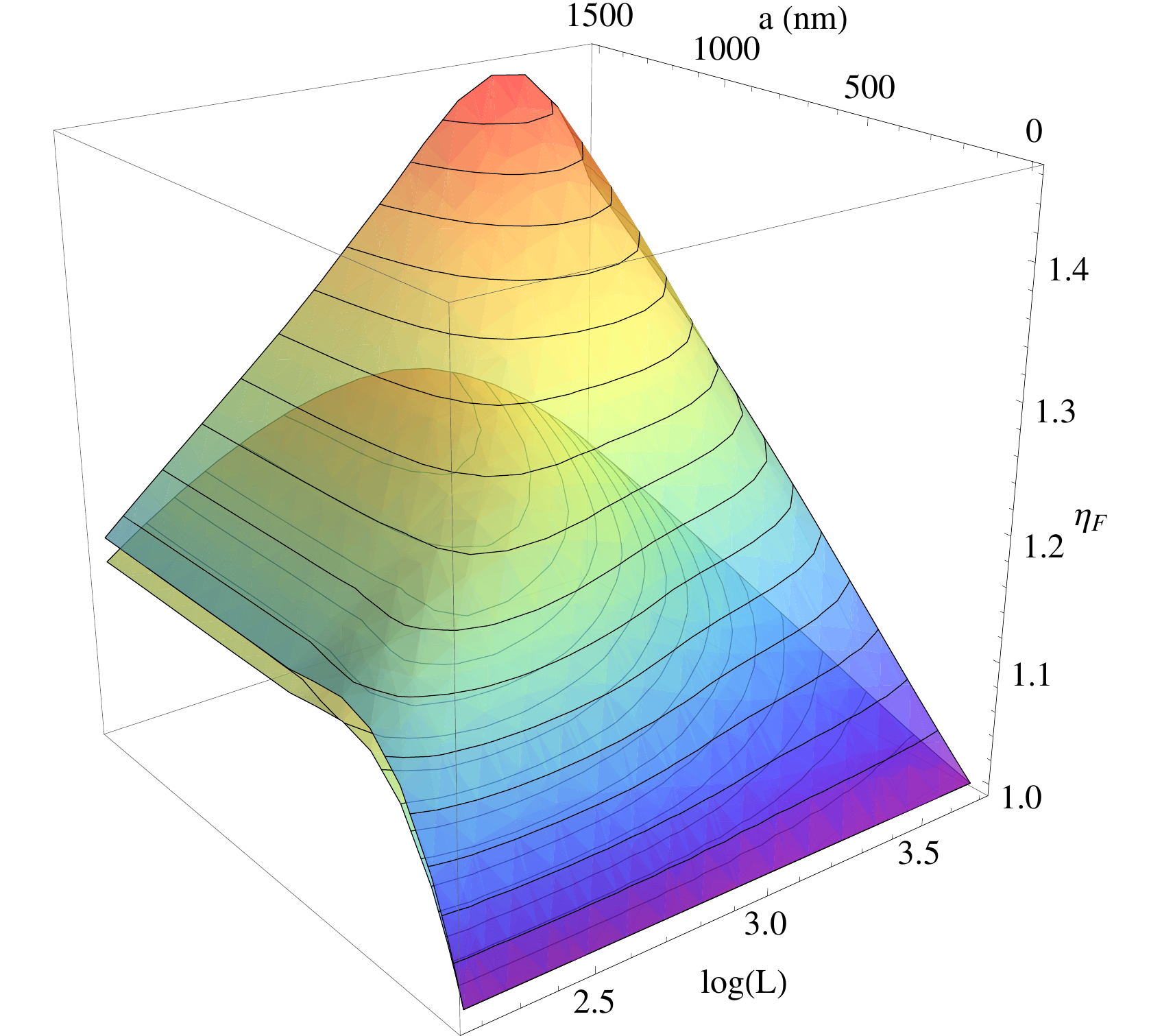}
  \end{center}
  \caption{Deviation from PFA $\eta_{F}$ as a function of the separation distance $L$ and trench depth $a$. The lower surface is for $T=0$ K, the upper one for $T=300$ K.}
  \label{fig:etaF}
\end{figure}

\begin{figure}[htbp]
  \begin{center}
    \includegraphics[width=3in]{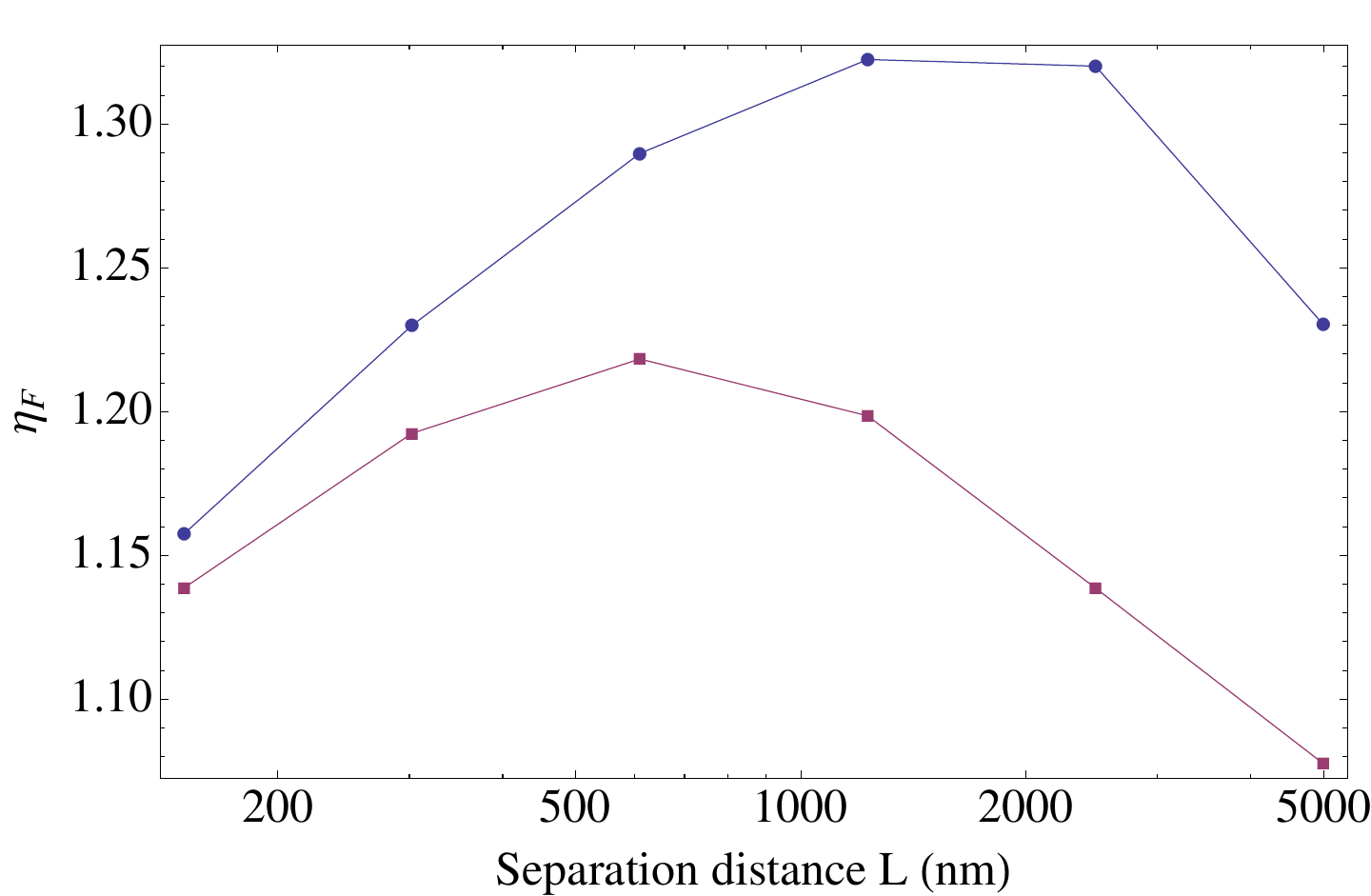}
  \end{center}
  \caption{Deviation from PFA $\eta_{F}$  at fixed trench depth $a=900$ nm as a function of the separation distance $L$ for $T=0$ K (bottom) and $T=300$ K (top).}
  \label{fig:fig3Cut_L_a900}
\end{figure}

Next, we turn to assess the validity of PFA. Figure~\ref{fig:etaF}
shows the deviation from PFA $\eta_{F}$ as a function of the
separation distance $L$ and the corrugation depth $a$. By definition
$\eta_{F}(L,a=0)=1$. For a fixed separation distance $L$, the error
made by PFA increases with deeper trench depth $a$.
We have $\eta_{F}(L,a;T=300\,K)>\eta_{F}(L,a;T=0\,K)$ for all values
of separation distance $L$ and trench depth $a$ so that a finite
temperature is seen to always increase the deviation from PFA. At
large separation distances, we expect PFA to be valid so that
$\lim_{L\to \infty}\eta_{F}(L,a;T)=1$. In particular, for a
fixed corrugation depth $a$, the functions $\eta_{F}$ show a maximum
for a particular distance $L=L_\mathrm{max}$ as illustrated in
figure~\ref{fig:fig3Cut_L_a900}.
\emph{Qualitatively}, we can say that this value
$L_\mathrm{max}\approx a$. More precisely we find
$L_{max}\lessapprox a$ for $T=0$ K and $L_{max}\gtrapprox a$ for
$T=300$ K. Thus the deviation from PFA increases with increasing
distances for $L \le a$ and decreases with increasing distances for
$L \ge a$. The deviation from PFA for $T\neq 0$ K shows the same
qualitative behavior as for $T=0$ K.

\subsubsection{Comparison with experimental data}
\label{ssub:comparison} We are now in the position to compare
our calculations with experimental data from
\cite{chanPRL2008,baoPRL2010}. In these experiments the interaction 
between a nanostructured silicon surface and a gold-coated sphere 
with a radius of 50 $\mu$m was measured. The force was detected at 
ambient temperature using a silicon micromechanical resonator onto 
which the gold sphere was attached. As the distance between the 
nanostructured surface and the sphere was varied, the change in 
the resonant frequency of the resonator was recorded. 
This quantity is proportional to the Casimir
force gradient $\partial_{L}F_{s,g}(L)$ between the gold sphere and
the silicon grating. Since the separation distance between the sphere
and the grating is small compared to the radius of the sphere,
we can relate this Casimir force gradient to the Casimir pressure
$F_{p,g}(L)$ between a plate and the grating as
$\partial_{L}F_{s,g}(L)=2\pi R F_{p,g}(L)$. Under these assumptions,
the measured quantity $\partial_{L}F_{s,g}(L)$ when normalized by
its PFA value is identical to $\eta_{F}$, i.e.
$\partial_{L}F(L)/\partial_{L}F^{PFA}(L)=\eta_{F}(L)$ where we
have omitted for simplicity the indices indicating the sphere
grating geometry.

\begin{figure}[htbp]
  \begin{center}
    \includegraphics[width=3in]{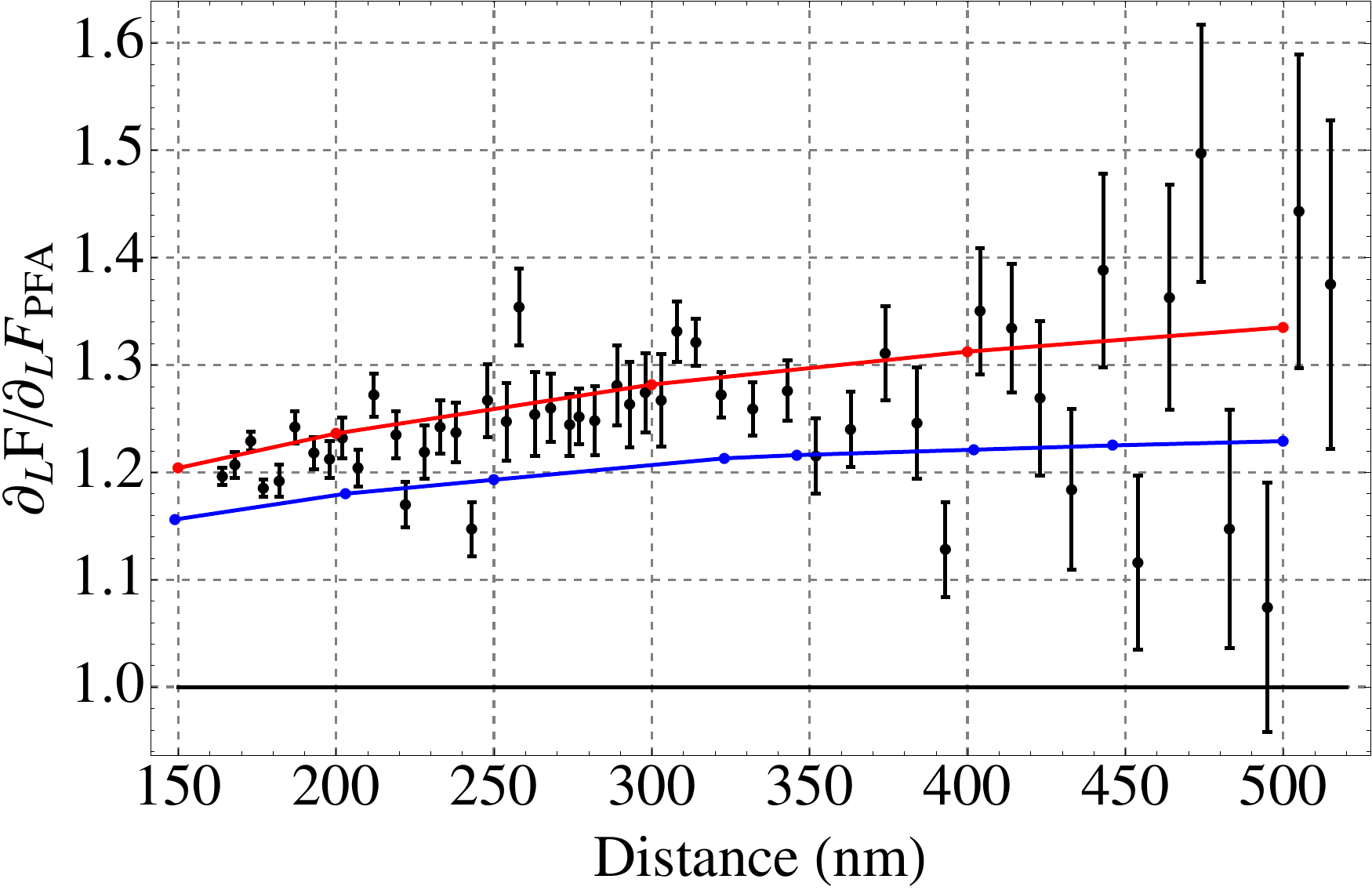}
  \end{center}
  \caption{Experimental data for the Casimir force gradient between a gold sphere and a silicon grating with deep trenches (black dots with error bars).
  The data are normalized to their PFA expression. This is compared with our calculations both for $T=0$ K (blue) and $T=300$ K (red).}
  \label{fig:deepChan}
\end{figure}

In figure~\ref{fig:deepChan} 
we plot $\eta_{F}(L)$ for a silicon structure with deep trenches
(sample B in Ref.~\cite{chanPRL2008}). This sample has a period
$d=400$ nm, trench depth $a=980$ nm and filling factor $f=0.478$.
The experimental data are plotted as circles with error bars.
Our calculations for $T=0$ K and $T=300$ K are plotted as
the blue and red curves respectively. As first reported in 
\cite{chanPRL2008}, the measured values of $\eta_{F}(L)$ between 
gold and silicon surfaces were found to be smaller than 
predictions using perfect reflectors at zero temperature \cite{buscherPRA2004}.
The inclusion of the material properties at zero temperature leads to better
agreement \cite{lambrechtPRL2008}. Our zero-temperature
results presented here are larger than those in
\cite{lambrechtPRL2008} by about $\partial_{L}F/\partial_{L}F^{PFA}
\approx 0.05$. We attribute this difference to the
replacement of the matrix $\boldsymbol{\epsilon}$ by
$\pmb{\backepsilon}^{-1}$ in the lower-left block of the matrix
$\mathbf{M}$ in Eq. (\ref{eq:matrixM2}) and to solving the
differential equations (\ref{eq:helm}) instead of
(\ref{eq:matrixM1}).

Let us now discuss the thermal effects. The red line in 
figure~\ref{fig:deepChan} plots the calculated results for $T=300$ K, 
at separation $L$ from $\approx 150$ nm to $L\approx 500$ nm.
Following our previous discussion of Fig.~(\ref{fig:fig3Cut_L_a900}), 
since the experiment was conducted in a regime where $L<a$, both
$\eta_F$ and $\partial_{L}F/\partial_{L}F^{PFA}$ increase
with separation distance. In figure~\ref{fig:deepChan}, the difference between the red line for 
$T=300$ K and the blue line for $T=0$ K is clearly visible.  
When compared to the experimental data measured at ambient temperature, 
the theory curve calculated for $T=300$ K gives better agreement
than the one at $T=0$ K. Unambiguous demonstration of the thermal 
contributions of the Casimir force in this sample, however, 
would require experimental improvements to further reduce the 
measurement uncertainty. 

So far, the thermal contributions to the 
Casimir force has only been observed at distances larger than $1 \mu$m between 
smooth surfaces~\cite{sushkovNatPh2011}. At smaller distances, the thermal
effects decrease significantly. As shown in figure~\ref{fig:thetaF_cut}, 
the thermal contributions to the Casimir force 
between flat surfaces are expected to be only about $3 \%$ at $\sim 500$ nm. By replacing 
one of the surfaces with a grating with deep trenches, the thermal contributions 
increase by a factor of 3. Nanostructured surfaces therefore hold promise for 
precise measurements of the thermal Casimir force.

\begin{figure}[htbp]
  \begin{center}
    \includegraphics[width=3in]{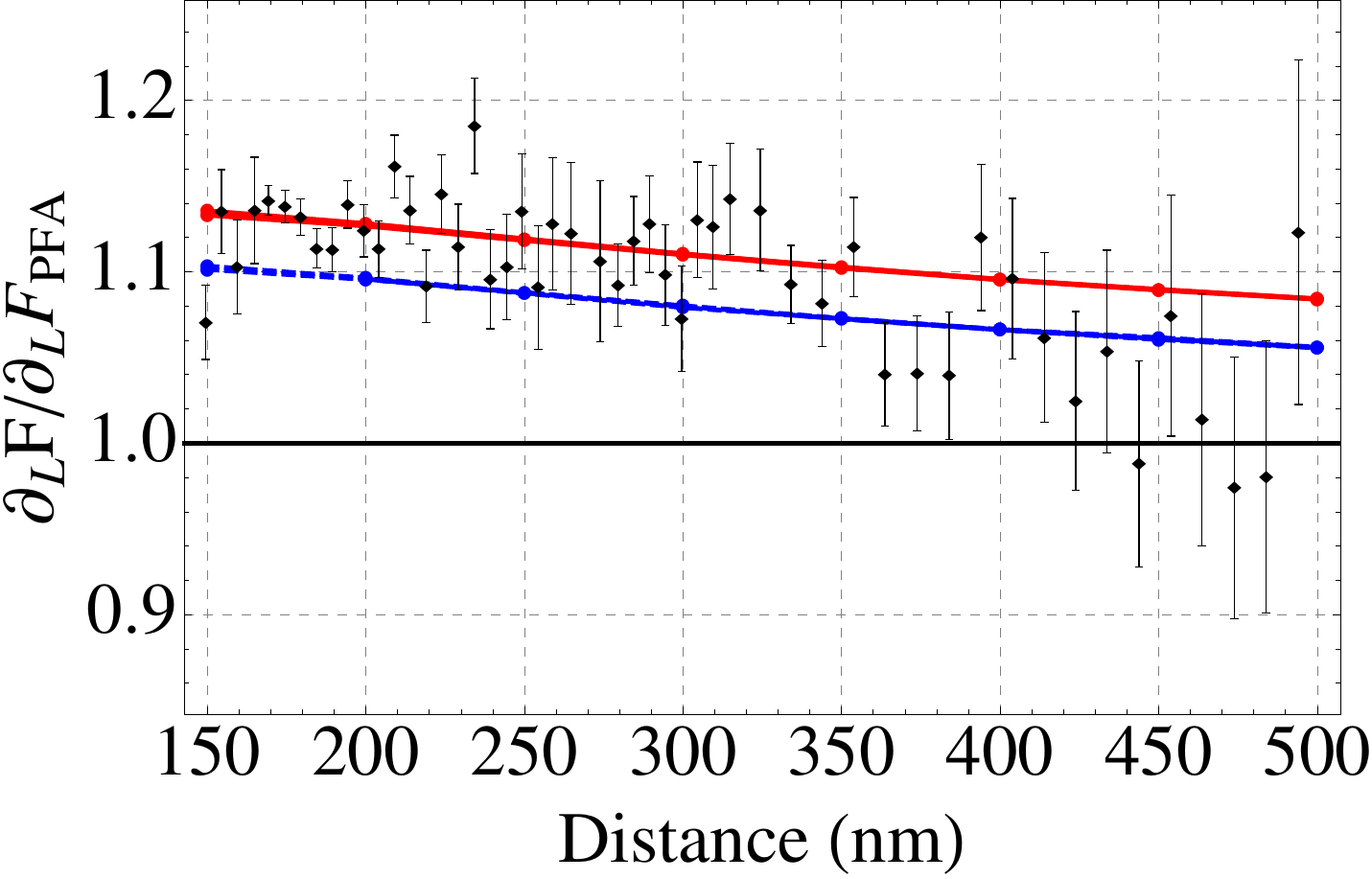}
  \end{center}
  \caption{Experimental data for the Casimir force gradient between a gold sphere and a silicon grating with shallow trenches (black dots with error bars).
  The data are normalized to their PFA expression. This is compared with our calculations both for $T=0$ K (blue) and $T=300$ K (red).}
  \label{fig:shallowChan}
\end{figure}

Next, we focus on gratings with shallow trenches that were  used in
\cite{baoPRL2010}. The period was again $d=400$ nm, but the trench depth 
was only $a=98$ nm. The filling factor was approximately $f=0.48$.
Figure~\ref{fig:shallowChan} shows the measured data points from
this experiment together with the results of our calculations for
this sample both at $T=300$ K (red curve) and $T=0$ K
(blue curve). In the calculation we take into account the exact
trapezoidal shape of the corrugation profile via a generalization of
the formalism presented above~\cite{lussangeCA2012}. As the range of
separation distances is the same as in the experiment with deep
trenches the situation now corresponds to a regime where $L>a$.
Therefore $\eta_F$ and thus $\partial_{L}F/\partial_{L}F^{PFA}$
both decrease with increasing distance to reach its asymptotic value of
1. Again the theoretical prediction at 300K is
in good agreement with the measured data. The overall temperature
effect is less pronounced here than for the deep trenches as the
trench depth is about a factor of 10 smaller.

\subsubsection{Conclusions}
\label{ssub:conclusions}

  We have calculated the Casimir interaction between a plate and a grating at finite temperature.
  We find good agreement between our calculations for $T=300$ K and experimental data taken at ambient temperature. Even though the experiments are performed
  at relatively small separation distances $L<500$ nm, the use of gratings enhances the thermal contributions of the Casimir force. Our findings provide an alternative approach to study thermal Casimir forces
  without having to reach separation distances of the order of microns.

H.B.C. is supported by HKUST 600511 from the Research Grants Council 
of Hong Kong SAR, Shun Hing Solid State Clusters Lab and DOE Grant No. DE-FG02-05ER46247.

We thank the European Science Foundation (ESF) within
the activity \emph{New Trends and Applications of the Casimir Effect}
(www.casimir-network.com) for support.

\providecommand{\noopsort}[1]{}\providecommand{\singleletter}[1]{#1}%

\end{document}